\documentclass[%
reprint,
superscriptaddress,
amsmath,amssymb,aps,pra]{revtex4-1}

\usepackage{graphicx}
\usepackage{dcolumn}
\usepackage{bm}
\usepackage{nicefrac}

\newcommand{\rme}{\mathrm{e}}
\newcommand{\inta}[1]{\int_0^{\infty}\mathrm{d}#1\ }

\newcommand{\mb}{\mathbf}
\newcommand{\ket}[1]{\left|#1\right>}

\newcommand{\p}{^{\prime}}

\newcommand{\uk}{_{\mathbf{k}}}

\newcommand{\uq}{_{\mathbf{q}}}
\newcommand{\uqp}{_{\mathbf{q}^{\prime}}}

\newcommand{\abs}[1]{\left|#1\right|}

\usepackage{etoolbox}
\preto\subequations{\ifhmode\unskip\fi} 

\usepackage[usenames,dvipsnames]{xcolor}

\begin{document}
	
	\preprint{APS/123-QED}
	
	\title{An analytic expression for the optical exciton transition rates in the polaron frame}
	
	\author{Dominic M. Rouse}
	\affiliation{SUPA, School of Physics and Astronomy, University of St Andrews, St Andrews, KY16 9SS, UK}
	
	\author{Erik M. Gauger}
	\affiliation{SUPA, Institute of Photonics and Quantum Sciences, Heriot-Watt University, Edinburgh, EH14 4AS, UK}
	
	\author{Brendon W. Lovett}
	\email{bwl4@st-andrews.ac.uk}
	\affiliation{SUPA, School of Physics and Astronomy, University of St Andrews, St Andrews, KY16 9SS, UK}
	
	\date{\today}%
	
	\begin{abstract}
	 When an optical emitter is strongly coupled to a vibrational bath the polaron transformation is often used to permit an accurate second-order Redfield master equation. However, the optical transition rates in the polaron frame are not analytic and approximations typically need to be made which result in the loss of  anything other than simple additive effects of the two baths. In this paper, we derive an intuitive analytic expression for the polaron frame optical transition rates by means of a finite mode truncation of the vibrational bath. Using this technique, calculations of the transition rates converge for only a few modes in the truncated spectral density, and capture non-additive effects such as population inversion of a two-level system.  
	\end{abstract}
	
	\maketitle
	
	\section{Introduction}
	Models of excitons interacting with both optical and vibrational baths are important in understanding a variety of physical processes, such as excitonic energy transfer \cite{kreisbeck2011high,ye2012excitonic,fruchtman2016photocell,hu2018dark,killoran2015enhancing,gelbwaser2017thermodynamic,creatore2013efficient,dorfman2013photosynthetic,wertnik2018optimizing,tomasi2021environmentally,tomasi2020classification,tomasi2019coherent} and superabsorption and superradiance \cite{dicke1954coherence,del2015quantum,higgins2014superabsorption,brown2019light,higgins2017quantum}. In these systems, the vibrational interaction originates from the vibration of ions surrounding the optically active site, for example in the solid-state lattice of a quantum dot \cite{nazir2016modelling,gies2007semiconductor,weig2004single,maier2011charge,kok2010introduction} or the protein scaffolding of an organic molecule \cite{clear2020phonon,del2015quantum,arnardottir2020multimode}. The optical interaction leads to the creation and annihilation of exciton states, whilst the vibrational interaction causes decoherence of these states. The decoherence plays the crucial role of permitting excitations to move between eigenstates of the exciton system, giving rise to energy transfer. This distinction of the roles of the optical and vibrational baths is only valid in the non-additive limit of simultaneously weak coupling to both baths. In the strong vibrational coupling regime, it has recently been shown that the presence of strongly coupled higher energy vibrational states can significantly renormalise optical transition rates \cite{maguire2019environmental,gribben2021exact}.
	
	A commonly used method to derive master equations in the strong vibrational coupling regime is to use the polaron transformation \cite{denning2020optical,bundgaard2021non,mccutcheon2011general,nazir2016modelling,kok2010introduction,qin2017effects,pollock2013multi,rouse2019optimal,tomasi2021environmentally,clear2020phonon,gribben2021exact,mccutcheon2011general,scerri2017method,nazir2009correlation,hughes2011influence,hughes2021resonant,restrepo2016driven,manson2016polaron,roy2011phonon,gustin2017influence}. A two-level exciton system coupled to only a vibrational bath that is displaced depending on the system eigenstate can be exactly diagonalised using the polaron transformation \cite{mahan2013many,nazir2016modelling,xu2016non}---this is the independent boson model. The diagonal basis describes polarons, which are quasiparticles consisting of the exciton and the phonons that are created when the vibrational bath is displaced. When the two-level system also interacts with an optical bath, the polaron transformation no longer diagonalises the Hamiltonian. Nonetheless, the description of the Hamiltonian in terms of polarons still allows one to contain the vibrational energy associated with the displacement of the bath within the unperturbed system Hamiltonian. Thus, a second-order master equation derived in the polaron frame will not break down due to strong vibrational coupling so long as the system is driven weakly \cite{nazir2016modelling,mccutcheon2011general}. The caveat is that the optical transition rates in the polaron frame are difficult to solve analytically owing to the non-additive interaction, and are numerically tractable only in special cases \cite{nazir2016modelling,pollock2013multi,qin2017effects,scerri2017method}. 
	
	Typically, expressions for the polaron frame optical transitions rates are found by employing approximations such as the flat spectral density approximation, where the optical spectral density is assumed constant in energy \cite{carmichael2009statistical,qin2017effects,nazir2016modelling,scerri2017method}. This results in the loss of non-additive effects because transitions between any two vibrational levels have the same weighting. In this paper we derive an intuitive analytic expression for the optical transition rates by approximating the vibrational bath as a finite number of modes. The expression converges and becomes numerically exact as the number of modes increases, although in many cases of interest a single mode approximation is accurate. Moreover, because this technique identifies a few-mode description of the bath, one can more easily interpret characteristic energy scales and coupling strengths.
	
	This paper is organised as follows: in Section~\ref{sec:Model} we introduce the simplest Hamiltonian where the polaron frame optical transition rates arise. In Section~\ref{sec:Polaron}, we transform to the polaron frame and, in Section~\ref{sec:Derivation}, we derive the general expression for the optical transition rates in terms of a \textit{polaron rate function}. This function is encountered whenever an optical transition rate is derived in the polaron frame even for more complicated exciton systems \cite{rouse2019optimal}. We then derive an analytic form of the polaron rate function which is the main result of the paper. Finally, in Section~\ref{sec:Example}, we benchmark the analytic form by comparing it to a numerical calculation in a case where such a calculation is possible. We then use the analytic method in cases where numerical integration is difficult.

	\section{Model}\label{sec:Model}
\begin{figure}[ht!]\centering
		\includegraphics[width=0.3\textwidth]{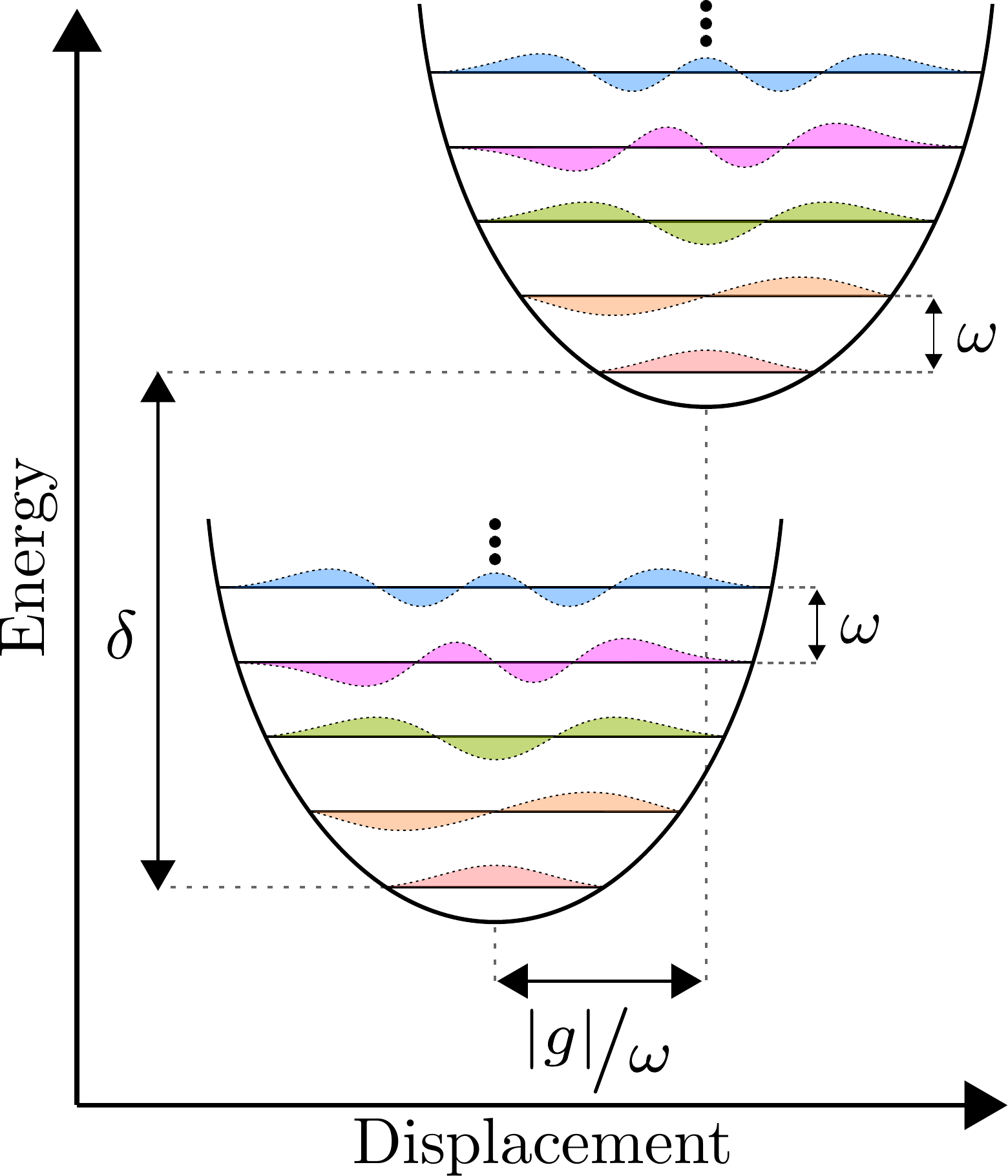}
		\caption{Schematic showing the energy levels of the vibronic Hamiltonian, $H_S+H_V$, in the case of a single vibrational mode with energy $\omega$ and coupling strength $g$.} \label{fig:vibexciton}
	\end{figure}

	The simplest model demonstrating the role of the polaron rate function (PRF) is a single two-level emitter coupled to both an optical and vibrational bath \cite{nazir2016modelling,xu2016non}. The optical interaction leads to transitions between the electronic states of the emitter by creation or annihilation of an exciton whilst the vibrational coupling is manifest in a conditional displacement of the nuclei making up the emitter, depending on the system state. The Hamiltonian describing this system can be partitioned into the exciton system (S), vibrational (V) and optical (O) parts as $H=H_S+H_V+H_O$. The exciton system part,
	\begin{align}\label{eq:HS}
		H_S=\delta\sigma^+\sigma^-,
	\end{align}
	describes the two electronic states of the emitter with transition energy $\delta$, where $\sigma^+$ takes the exciton from the ground state $\ket{g}$ to the excited state $\ket{e}$ of the two-level emitter, and $\sigma^- = (\sigma^+)^\dag$. The vibrational part has two contributions,
	\begin{equation}\label{eq:HV}
	H_V=\sum\uk\omega\uk b\uk^{\dagger}b\uk+\sigma^+\sigma^-\sum\uk\left(g\uk b\uk^{\dagger}+g\uk^*b\uk\right),
	\end{equation}
	where the first term describes the energy of the bath and the second term the conditional displacement of the nuclei. The vibrational bath is composed of phonons of wavenumber $\mb{k}$ and energy $\omega\uk$, characterised by the ladder operators $b\uk$. Each of these modes couples to the emitter with strength $g\uk$. This kind of interaction leads to Franck-Condon physics  \cite{kok2010introduction,maguire2019environmental}. When the vibrational coupling is strong, it is natural to consider $H_S+H_V$ as the unperturbed energy basis, and Figure~\ref{fig:vibexciton} shows the energy levels for this in the case of a single vibrational mode.
	
	The optical contribution to the Hamiltonian also contains a bath energy and an interaction term. This bath is composed of photons characterised by the ladder operators $a\uq$ with energy $\nu\uq$ where $\mb{q}$ is the wavenumber of the photon. Each mode couples to the emitter with strength $f\uq$. Within the electric dipole approximation the optical part is
	\begin{equation}
	H_O=\sum\uq\nu\uq a\uq^{\dagger}a\uq+\sigma^x\sum\uq \left(f\uq a\uq^{\dagger}+f\uq^* a\uq\right).\label{eq:HO}
	\end{equation}
	The interaction term describes the absorption or emission of a photon along with the creation or annihilation of an exciton in the emitter, including processes that do not conserve particle number.
	
	The coupling strengths to the baths are defined by the spectral densities $J_V(\omega)=\sum\uk\left|g\uk\right|^2\delta(\omega-\omega\uk)$ and $J_O(\nu)=\sum\uq\left|f\uq\right|^2\delta(\nu-\nu\uq)$ \cite{breuer2002theory,nazir2009correlation}. The continuum form of the vibrational spectral density is determined by the physical nature of the emitter, for example whether it is a quantum dot or a chromophore \cite{nazir2016modelling,sowa2018beyond,du2018theory,da2006zero,konig1996zero,dong1986molecular}. 
	The optical spectral density is more subtle, depending on the gauge in which the Hamiltonian is derived \cite{stokes2012extending,stokes2018master,rouse2021avoiding}.

	\section{Polaron transformation}\label{sec:Polaron}
	The polaron transformed Hamiltonian is $H\p=U^\dagger HU$ where the unitary operator is $U=\mathrm{exp}\left[-G\sigma^+\sigma^-\right]$ with
	\begin{equation}
	G=\sum\uk\left[\frac{g\uk}{\omega\uk}b\uk^{\dagger}-\left(\frac{g\uk}{\omega\uk}\right)^*b\uk\right].
	\end{equation}
	This can be rewritten in the more revealing form
	\begin{equation}\label{eq:UP}
	U=\sigma^-\sigma^++B\sigma^+\sigma^-,
	\end{equation}
	where $B=\exp\left(-G\right)$ is a displacement operator. The displacement operator transforms the vibrational ladder operators as $B^\dagger b\uk B=b\uk-g\uk/\omega\uk$,	and its name is attributed to this property. Eq.~\eqref{eq:UP} shows that the polaron transformation leaves the ground electronic state unaffected but displaces each vibrational mode coupled to the excited electronic state by $g\uk/\omega\uk$. 
	
	The polaron transformation exactly diagonalises the independent boson model, i.e. our model without $H_O$:
	\begin{equation}
	U^\dagger(H_S+H_V)U=\delta^{\prime}\sigma^+\sigma^-+\sum\uk\omega\uk b^{\dagger}\uk b\uk,
	\end{equation} 
	where $\delta^{\prime}=\delta-\lambda$ is the polaron energy and 
	\begin{equation}\label{eq:reorg}
	\lambda=\sum\uk\frac{\abs{g\uk}^2}{\omega\uk}=\inta{\omega} \frac{J_V(\omega)}{\omega},
	\end{equation}
	is the reorganisation energy \cite{mahan2013many}. The vibrational interaction has been removed in place of renormalised electronic energy levels. Incorporating again the optical part, the full polaron frame Hamiltonian is  
	\begin{align}\label{eq:HP}
	H\p=\ &\delta^{\prime}\sigma^+\sigma^-+\sum\uk\omega\uk b^{\dagger}\uk b\uk+\sum\uq\nu\uq a\uq^{\dagger}a\uq\nonumber\\
	&+\left(B^{\dagger}\sigma^++B\sigma^-\right)\sum\uq \left(f\uq a^{\dagger}\uq+f^*\uq a\uq\right).
	\end{align}
	  Unlike in the independent boson model, the polaron transformation has not completely removed the  vibrational interaction. Whilst the original vibrational interaction in Eq.~\eqref{eq:HP} has been removed, there are now displacement operators appearing in the optical interaction. This arises since  the upper electronic manifold has been displaced by the vibrational interaction relative to the lower manifold; optical transitions either start or end in the displaced manifold, creating ($B^{\dagger}\sigma^+$) or annihilating ($B\sigma^-$) polarons of energy $\delta\p$. The polaron transformation is particularly useful because diagonal elements of the density matrix commute with the transformation, so polaron and exciton populations are equivalent.
	 
	 \section{Optical transition rates}\label{sec:Derivation}
	 The second order Born-Markov master equation in the polaron frame can be derived in the usual way \cite{breuer2002theory}. One finds that the excited state population evolves as
	 \begin{equation}\label{eq:rhoeePol}
	 \dot{\rho}_{ee}(t)=\gamma_{\uparrow}\rho_{gg}(t)-\gamma_{\downarrow}\rho_{ee}(t),
	 \end{equation}
	 and the ground state evolves as $\dot{\rho}_{gg}=-\dot{\rho}_{ee}$ where $\gamma_{\uparrow}$ and $\gamma_{\downarrow}$ are the excitation and decay rates, respectively. The population master equations are naturally decoupled from the coherences in this model; we have not made the secular approximation. The rates are
	 \begin{subequations}\label{eq:garrow}
	 	\begin{align}
	 	&\gamma_{\uparrow}=\gamma(-\delta\p),\label{eq:gexcite}\\
	 	&\gamma_{\downarrow}=\gamma(\delta\p),\label{eq:gdecay}
	 	\end{align}
	 \end{subequations}
 written in terms of the PRF, which has the form:
	 \begin{align}\label{eq:PRF}
	 \gamma(\eta)=2\mathrm{Re}&\inta{t}\Bigg(\rme^{i\eta t}\mathrm{Tr}_V\left[B^{\dagger}(t)B(0)\rho_V\right]\nonumber\\
	 &\hspace{1cm}\times\mathrm{Tr}_O\left[\sum_{\mathbf{q}\mathbf{q}\p}A\uq^{\dagger}(t)A\uqp(0)\rho_O\right]\Bigg),
	 \end{align}
	 where $\rho_V$ and $\rho_O$ are the Gibbs states of the vibrational and optical baths, $B(t)$ is the displacement operator in the interaction picture and $A\uq(t)$ is the interaction picture form of the operator $A\uq=f\uq a\uq^{\dagger}+f\uq^*a\uq$. 
	 
	 Integrals of the form of Eq.~\eqref{eq:PRF} appear when studying optical interactions in the polaron frame. Currently, there is no analytic expression for these, and numerical solutions only exist for specific vibrational spectral densities. We will find an analytic expression. 
	 
	 In Eq.~\eqref{eq:PRF}, the trace over displacement operators is
	 \begin{equation}\label{eq:trdisp}
	 \mathrm{Tr}_V\left[B^{\dagger}(t)B(0)\rho_V\right]=\rme^{\phi(t)-\phi(0)},
	 \end{equation}
	 where, taking the continuum limit of $\mb{k}$, the phonon propagator is defined as
	 \begin{align}\label{eq:phi}
	 \phi(t)=\inta{\omega} \frac{J_V(\omega)}{\omega^2}\Big[&\cos\left(\omega t\right)\coth\left(\frac{\beta_V\omega}{2}\right)\nonumber\\
	 &\hspace{1.9cm}-i\sin\left(\omega t\right)\Big],
	 \end{align}
	 and $\beta_V=1/(k_BT_V)$ with $T_V$ being the temperature of the vibrational bath and $k_B$ the Boltzmann constant \cite{nazir2016modelling}. The trace over the optical operators is
	 \begin{align}
	 &\mathrm{Tr}_O\left[A\uq(t) A\uqp(0)\rho_O\right]\\\nonumber&=\delta_{\mb{q}\mb{q}\p}\Big(f\uq^*f\uqp\left[N_O(\nu\uq)+1\right]\rme^{-i\nu\uq t}+f\uq f\uqp^*N_O(\nu\uq)\rme^{i\nu\uq t}\Big),
	 \end{align}
	 where $N_O(\nu)=1/\left(\rme^{\beta_O\nu}-1\right)$ is the population of the photon mode with energy $\nu$ at the temperature of the optical bath $T_O=1/(k_B\beta_O)$. After taking the continuum limit of the optical wavenumber $\mb{q}$, the PRF becomes
	 \begin{align}\label{eq:PRF2}
	 \gamma(\eta)=\inta{\nu}\Big[ J_E(\nu)\mathcal{K}(\eta-\nu)+J_A(\nu)\mathcal{K}(\eta+\nu)\Big],
	 \end{align}
	 where we have defined the emission and absorption optical spectral densities,
	 \begin{subequations}\label{eq:JAJE}
	 	\begin{align}
	 	&J_E(\nu)=2\pi J_O(\nu)\left[1+N_O(\nu)\right],\label{eq:JE}\\
	 	&J_A(\nu)=2\pi J_O(\nu)N_O(\nu),\label{eq:JA}
	 	\end{align} 
	 \end{subequations}
	 and
	 \begin{equation}\label{eq:K}
	 \mathcal{K}(\epsilon)=\frac{1}{\pi}\mathrm{Re}\inta{t}\rme^{\phi(t)-\phi(0)}\rme^{i\epsilon t},
	 \end{equation}
	 which contains all of the effects of the vibrational coupling. The PRF in Eq.~\eqref{eq:PRF2} describes the non-additive physics of the vibrational and optical interactions. That the effects of these environments are non-additive is evident because Eq.~\eqref{eq:PRF2} is a convolution of purely optical and vibrational functions.
	 
	 Before we go onto derive the analytic expression to Eq.~\eqref{eq:PRF2} we will consider two common approximations used to obtain analytic expressions.
	 
	 \subsection{Approximate expressions}
	 The first approximation is the weak vibrational coupling limit. This is defined by $\phi(t)\to 0$, leading to 
	 \begin{equation}\label{eq:Kweak}
	 \mathcal{K}^{\mathrm{weak}}(\epsilon)=\delta(\epsilon),
	 \end{equation}
	 where we have used the identity
	 \begin{equation}\label{eq:CPidentity}
	 \mathrm{Re}\inta{t}\rme^{i\alpha t}=\pi\delta(\alpha),
	 \end{equation}
	 valid only within another integral \cite{breuer2002theory}. Inserting this into Eq.~\eqref{eq:PRF2} results in
	 \begin{subequations}\label{eq:gweak}
	 	\begin{align}
	 	\gamma^{\mathrm{weak}}_{\uparrow}= J_A(\delta),\\
	 	\gamma^{\mathrm{weak}}_{\downarrow}= J_E(\delta).
	 	\end{align}
	 \end{subequations}
	 Therefore, the excitation and decay rates will only sample the optical spectral density at the electronic splitting $\delta$, corresponding to a transition between the two ground vibrational levels in each manifold. 
	 
	 The second typical approximation is the flat spectral density approximation. In this approximation the optical spectral density is assumed to be constant around the polaron energy, i.e. $J_A(\epsilon>0)=J_A(\delta\p)$ and $J_E(\epsilon>0)=J_E(\delta\p)$ and both are zero for $\epsilon\le 0$. In this case, the frequency integral in Eq.~\eqref{eq:PRF2} can be performed before the time integral in Eq.~\eqref{eq:K}. Then using
	 \begin{equation}\label{eq:Kflat}
	 \inta{\nu}\mathcal{K}(\eta\pm\nu)=1,
	 \end{equation}
	leads to
	 \begin{subequations}\label{eq:gflat}
	 	\begin{align}
	 	\gamma^{\mathrm{flat}}_{\uparrow}= J_A(\delta\p),\\
	 	\gamma^{\mathrm{flat}}_{\downarrow}=J_E(\delta\p).
	 	\end{align}
	 \end{subequations}
	 In this approximation, the vibrational coupling strength only enters through the polaron energy $\delta\p$. This approximation accounts for transitions between all vibrational levels, but because the optical spectral density is assumed to be flat each transition is given the same weighting. The sum over the the vibrational wavefunctions then involves only the overlap integrals between the vibrational eigenstates. Since these form a complete basis the vibrational contribution is completely removed. Mathematically, this occurs in Eq.~\eqref{eq:Kflat}. In both of these typical approximations, excitation and decay of the system only occur by photon absorption and emission, respectively.

	 \subsection{Analytic form of the PRF}
	 We will now derive an analytic form of the PRF, in Eq.~\eqref{eq:PRF2}, as a convergent series. The expression is valid for any vibrational spectral densities with weighted moments 
	 \begin{equation}\label{eq:weightedmo}
	 \mu_j=\inta{\omega}\frac{J_V(\omega)}{\omega^2}\omega^j,
	 \end{equation}
	 that are finite for $j=1,2,3,\ldots,\infty$. This is the same condition under which the polaron transformation does not lead to divergent expressions. To derive the series form we make use of two properties of the PRF which we will prove in the remainder of this subsection.
	 
	 1) The first property is that the PRF can be solved analytically for the truncated spectral density,
	 \begin{equation}\label{eq:JVeff}
	 J\p_V(\omega)=\sum_{i=1}^{N_*}\left|g\p_i\right|^2\delta\left(\omega-\omega\p_i\right),
	 \end{equation}
	 which is identical to $J_V$ except that the number of modes is truncated to $N_*$. As we will show, the form for this truncated spectral density is easily derived for a single vibrational mode $N_*=1$ which can then be extended to general $N_*$. 
	 
	 2) The second property is that $\mathcal{K}(\epsilon)$, in Eq.~\eqref{eq:K}, which contains the entire vibrational contribution to the PRF, is completely determined by the weighted moments of the vibrational spectral density, $\mu_j$. We will see that in general the orders $j$ of $\mu_j$ that contribute to the PRF are determined by the vibrational temperature $T_V$. As we will show, we can use this property to calculate the coupling strength and mode energy of each mode in the truncated spectral density for a chosen $N_*$.

	 \subsubsection{Proof of  property 1: analytic form of the PRF}
	 We will now prove the first property, that the PRF can be derived analytically for the truncated spectral density in Eq.~\eqref{eq:JVeff}. We will do this by first deriving the analytic expression for the PRF with a single mode, $N_*=1$, and then we will extend this to a general $N_*$ truncation.
	 
	 For $N_*=1$ in Eq.~\eqref{eq:JVeff}, the truncated phonon propagator can be written as a sum of zero and finite vibrational temperature parts
	 $\phi^s(t)=\phi^s_0(t)+\phi^s_{\beta}(t)$, where
	 \begin{subequations}\label{eq:phisinglesplit}
	 \begin{align}
	 &\phi^s_0(t)=S_1\p\rme^{-i\omega_1\p t},\\
	 &\phi^s_{\beta}(t)=S_1\p N_{\omega_1\p}\left(\rme^{i\omega_1\p t}+\rme^{-i\omega_1\p t}\right),
	 \end{align} 
	 \end{subequations}
	 and $N_{\omega_1\p}=1/(\rme^{\beta_V\omega_1\p}-1)$, $S_1\p=(\abs{g_1\p}/\omega_1\p)^2$ is the single mode Huang-Rhys parameter and the superscript `$s$' denotes a single vibrational mode. We then expand $\exp[{\phi^s(t)}]$ in Eq.~\eqref{eq:K} as a Taylor series and subsequently each $[\phi^s(t)]^n$ as a binomial series to find the single mode form of  Eq.~\eqref{eq:K}:
	 \begin{align}\label{eq:K2}
	 \mathcal{K}^s(\epsilon)=&\frac{1}{\pi}\rme^{-\phi^s(0)}\sum_{n=0}^{\infty}\sum_{m=0}^n\frac{1}{n!}{m \choose n}\nonumber\\
	 &\times\mathrm{Re}\inta{t}\rme^{i \epsilon t}[\phi_0^s(t)]^{n-m}[\phi_{\beta}^s(t)]^m,
	 \end{align}
	 where ${m\choose n}$ is the binomial coefficient. The indices $n$ and $m$ of each term in the sum of Eq.~\eqref{eq:K2} have physical meaning: $n$ denotes the total number of phonons contributing; $m$ denotes the order of finite $T_V$ contribution and $n-m$ denotes the $T_V=0$ contribution. Substituting in Eqs.~\eqref{eq:phisinglesplit} and making a final binomial expansion of $[\phi_\beta^s(t)]^m$ leads to
	 	\begin{align}	\mathcal{K}^s(\epsilon)&=\sum_{n=0}^{\infty}\sum_{m=0}^n\sum_{k=0}^m{n \choose m}{n\choose k}\label{eq:Ks2}\\
	 	&\hspace{0.5cm}\times W_n(S_1\p)V_m(S_1\p,\omega_1\p)\delta\left(\epsilon-\left[n-2m+2k\right]\omega_1\p\right),\nonumber
	 	\end{align}
	 where we have used Eq.~\eqref{eq:CPidentity} to take the real part of the integral \footnote{The imaginary part leads to Lamb shifts of the polaron frame splitting $\delta\p$. These are divergent unless a high frequency cut-off is imposed on the optical spectral density.}. We have defined $W_n(S)=S^n\exp[-S]/n!$ and the temperature factor $V_m(S,\omega)=N^m_\omega\exp[-2SN_\omega]$. Eq.~\eqref{eq:Ks2} shows that the PRF will be a sum over optical transitions between any two vibrational levels in the electronic manifolds. The probability for an $n$ phonon transition is $W_n$, and the probability for an $m^{\text{th}}$ order temperature transition is $V_m$. Specifically, $V_m$ is the probability that the initial and/or final state of a transition will be vibrational levels that are in total $m$ levels higher than accessible at $T_V=0$.
	 
	 Finally, we rewrite Eq.~\eqref{eq:Ks2} in a more compact form by grouping all transitions occuring at a common energy,
	 \begin{equation}
	 \mathcal{K}^s(\epsilon)=\sum_{l=-\infty}^\infty A_l(S_1\p,\omega\p_1)\delta(\epsilon-l\omega_1\p),
	 \end{equation}
	 where the amplitude coefficient is
	 \begin{equation}\label{eq:Al}
	 A_l(S,\omega)=\sum_{n=\left|l\right|}^{\infty+}\sum_{m=\frac{n-l}{2}}^n{n \choose m}{m \choose m-\frac{n-l}{2}}W_n(S)V_m(S,\omega),
	 \end{equation}
		and the `+' on the summation indicates that only every other term is included, i.e. $n=\left|l\right|,\left|l\right|+2,\ldots$. The amplitude coefficient describes the vibrational overlap and thermal occupation of the vibrational levels separated by $l$-levels. In practice, the sum over $n$ must be truncated: strong vibrational coupling means that more vibrational levels can be reached in a transition.
		
\begin{figure}[h!]\centering
	\vspace{0.5cm}
		\includegraphics[width=0.45\textwidth]{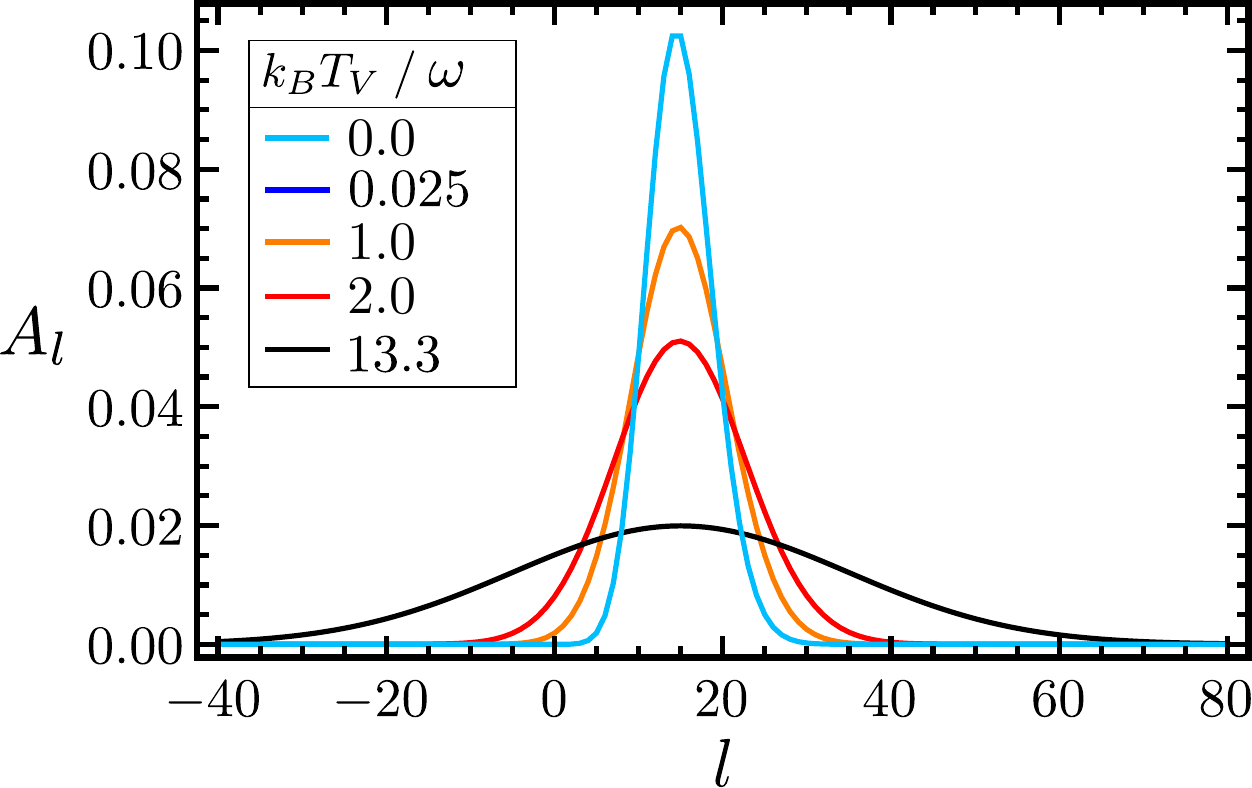}
			\caption{The amplitude coefficient $A_l(S,\omega)$ in Eq.~\eqref{eq:Al} is plotted as a function of $l$ and vibrational temperature $T_V$ for a phonon mode with $S=15$ and $\omega=1$~eV. At $T_V=0$ the distribution is exactly a Poisson distribution. At $T_V>0$ the distribution widens, flattens and extends into $l<0$. The $T_V=0$~K and $T_V=298$~K ($k_BT_V/\omega=0.025$) curves overlap.}\label{fig:Al}
		\end{figure}

	The distribution $A_l$ is maximised for $l=\text{round}(S)$ at all $T_V$, which means that the most likely optical transition will involve $\text{round}(S)$ phonons \footnote{$\text{round}(S)$ is the nearest integer to $S$.}. $A_l$ also has the normalisation property $\sum_lA_l(S,\omega)=1$, which is the same statement as Eq.~\eqref{eq:Kflat}. A well-known limit is $T_V=0$, in which case $V_0=1$ and $V_{m>0}=0$, and so $A_{l\ge 0}(S,\omega)=W_l(S)$ with $A_{l<0}(S,\omega)=0$ which is known as the Poisson Franck-Condon factor. In Figure~\ref{fig:Al} we plot $A_l$ as a function of $l$ and $T_V$. 
	
	For finite $T_V$, it can become slow to evaluate $A_l$ at strong vibrational coupling where the truncation of the sum over $n$ is necessarily high. However, in situations with strong vibrational coupling and low but finite $T_V$, many vibrational levels included in the sum over $n$ will not be thermally accessible, resulting in many of the terms in the sum over $m$ being negligible. Therefore, it is possible to replace
the upper limit of the sum over $m$ with $\text{Min}[n,\tilde{m}]$, where $\tilde{m}<n$ is determined by the size of $V_m$ as $m$ increases.

		Substituting $\mathcal{K}^s$ into Eq.~\eqref{eq:PRF2} leads to the single mode PRF,
		\begin{equation}\label{eq:PRFsingle}
		\gamma^s(\eta)=\sum_{l=-\infty}^{\infty}A_l(S\p_1,\omega\p_1)\left[J_E(\eta-l\omega\p_1)+J_A(-\eta+l\omega\p_1)\right].
		\end{equation}
		This intuitive result describes optical transitions between any two vibrational levels between the excited and ground manifolds. 
		
\begin{figure}[ht!]\centering
		\includegraphics[width=0.45\textwidth]{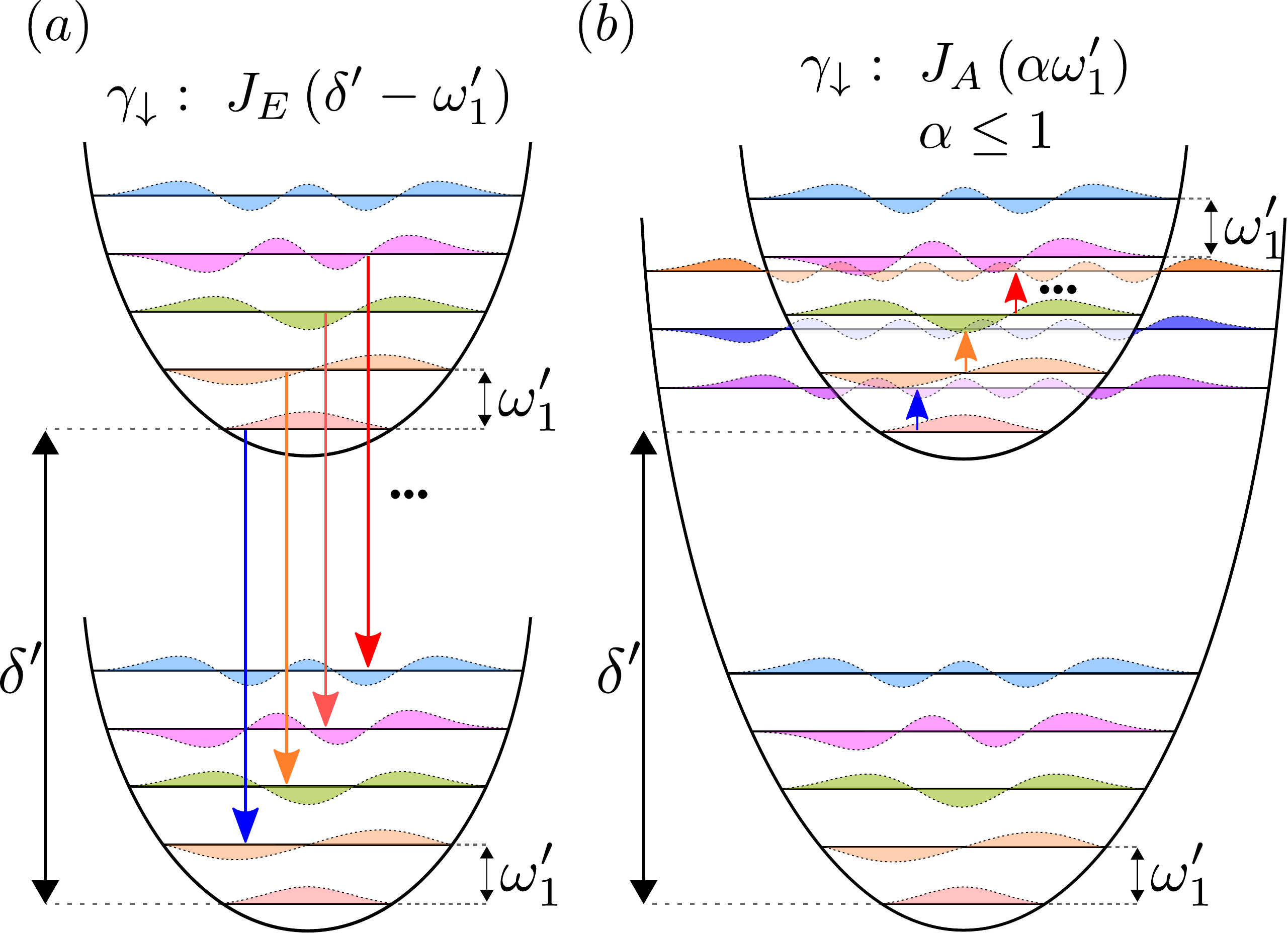}
		\caption{Exemplary decay transitions from the upper to the lower manifold by (a) photon emission and (b) photon absorption, drawn for $N_*=1$ in the truncated spectral density. We draw a series of transitions corresponding to the same frequency. These transitions (and all others of the same frequency) contribute to the corresponding $l$ in the $A_l$ distribution of Eq.~\eqref{eq:Al}. That is, in (a) the transitions contribute to $l=1$ and in (b) to $l=l\p\in\mathbb{Z}$ where $-\delta\p+l\p\omega=\alpha\omega$ for $0<\alpha\le1$. The colour of the transition arrow indicates how many finite temperature contributions there are to the corresponding term in $A_l$; blue represents $T_V=0$ ($m=0$) and approaching red means greater $T_V$ ($m>0$). Compared to Figure~\ref{fig:vibexciton} there is no relative displacement of the manifolds as this is in the polaron frame} \label{fig:viblevelsmain}
	\end{figure}

		Owing to the counter rotating terms in the optical interaction, system excitation can occur by photon emission and conversely system decay by photon absorption with the dominant process depending on the vibrational coupling strength and temperature. For example, let us consider decay processes. At weak coupling, the dominant decay channel is photon emission but this swaps to photon absorption at strong coupling. One can see this by noting that $\gamma_\downarrow^s=\gamma^s(\delta\p)$ and $A_l(S,\omega)$ is maximised for $l= \text{round}(S)$. Therefore, when $\delta\p-S_1\p\omega_1\p\gg0$, decay occurs via photon emission and because $J_O(\nu<0)=0$ there is no photon absorption. However, when $\delta\p-S_1\p\omega_1\p\ll0$ decay occurs exclusively through photon absorption. Exemplary decay transitions via both photon emission and absorption are shown in Figure~\ref{fig:viblevelsmain}.
		
		Following similar arguments one can show that excitation occurs only via photon absorption at all vibrational coupling strengths if $T_V=0$. This is because excitation by photon emission requires transitioning from a vibrational state in the ground manifold with energy $\epsilon>\delta\p$, requiring $k_BT_V>\delta\p$.

	 	For an arbitrary $N_*$ in the truncated spectral density, in Eq.~\eqref{eq:JVeff}, the phonon propagator is
	 	\begin{equation}
	 	\phi(t)=\sum_{i=1}^{N_*} S_i\p\left[\cos\left(\omega_i\p t\right)\coth\left(\frac{\beta_V\omega_i\p}{2}\right)-i\sin\left(\omega_i\p t\right)\right].
	 	\end{equation}
	 Repeating the same derivation as with the $N_*=1$ case, albeit with significantly more algebra, leads to the result that the full PRF for the truncated spectral density is
	 \begin{widetext}
	 \begin{equation}\label{eq:PRFNs}
	 \gamma(\eta) = \prod_{j=1}^{N_\ast}\sum_{l_j=-\infty}^{\infty} A_{l_j}(S_j\p,\omega\p_j)\left[J_E\left(\eta-\sum_{i=1}^{N_\ast}l_i\omega_i^\prime\right)\right.\left.+J_A\left(-\eta + \sum_{i=1}^{N_\ast}l_i\omega_i^\prime\right)
	 \right].
	 \end{equation}
	 \end{widetext}
	 One can see that Eq.~\eqref{eq:PRFNs} is the intuitive extension of the rate expressions in the single mode case in Eq.~\eqref{eq:PRFsingle} to allow for optical transitions from any combination of vibrational levels in the $N_*$ harmonic ladders to any other combination of levels in these ladders.
	 
	 It is interesting to note that the rate equations within the flat spectral density approximation in Eqs.~\eqref{eq:gflat} can be recovered from Eq.~\eqref{eq:PRFNs} by using the fact that under the flat approximation $J_E(\eta-\sum_{i=1}^{N_*}l_i\omega_i\p)=J_E(\eta)$ and $J_A(-\eta+\sum_{i=1}^{N_*}l_i\omega_i\p)=J_E(-\eta)$ and applying the property $\sum_lA_l(S,\omega)=1$. 
	 
	 Eq.~\eqref{eq:PRFNs} is the analytic expression for the PRF and is the main result of this paper. We will now prove the second property of the PRF, that the vibrational dependence of the PRF is determined by the weighted moments $\mu_j$ of the continuum spectral density $J_V$. In doing so we will derive a method to calculate the $2N_*$ parameters of the $N_*$ modes, $\{S_i\p\}$ and $\{\omega_i\p\}$, of the truncated spectral density $J_V\p$.

	 \subsubsection{Proof of property 2: truncated mode parameters}
	 
	 To prove the second property we will show that the PRF with the infinite mode spectral density that we wish to model is equivalent to the PRF with the truncated spectral density in Eq.~\eqref{eq:JVeff}, given in Eq.~\eqref{eq:PRFNs}. This proof depends on the vibrational bath temperature.
	 
	 At $T_V=0$ the phonon propagator we wish to model, in Eq.~\eqref{eq:phi}, is
	 \begin{equation}	     \phi(t)-\phi(0)=\inta{\omega}\frac{J_V(\omega)}{\omega^2}\left[\rme^{-i\omega t}-1\right].
	 \end{equation}
	 After expanding the factor of $\exp\left(-i\omega t\right)$ we obtain
	 \begin{equation}\label{eq:phiT0}
	\phi(t)-\phi(0)=\sum_{j=1}^\infty\frac{1}{j!}(-it)^j\mu_j.
	 \end{equation}
	 Substituting this into Eq.~\eqref{eq:K} and expanding the factor of $\exp(i\epsilon t)$ leads to
	 \begin{equation}\label{eq:ST0inf} 
	 \mathcal{K}(\epsilon)=\sum_{n=0}^{\infty}\frac{1}{n!}k_n(\epsilon),
	 \end{equation}
	 with $k_0(\epsilon)=\delta(\epsilon)$ and
	 \begin{equation} \label{eq:ST0inf2}
	 k_{n\ge 1}=\sum_{j_1,...,j_n=1}^{\infty}(-1)^{j_T}\frac{\mu_{j_1}\mu_{j_2}...\mu_{j_n}}{j_1!j_2!...j_n!}\delta^{(j_T)}(\epsilon),
	 \end{equation}
	 where $j_T=\sum_{i=1}^{n}j_i$ and $\delta^{(i)}(\epsilon)=(\partial^i/\partial \epsilon^i)\delta(\epsilon)$. Therefore, the only properties of the vibrational spectral density that determine the $T_V=0$ PRF are the moments $\mu_j$ for $j=1,2,3,\ldots,\infty$. Thus, the PRF derived using the truncated spectral density $J_V\p$ is identical to the desired PRF, calculated with $J_V$, if $\mu_j\p=\mu_j$ for $j=1,2,3,\ldots,\infty$ where 
	 \begin{equation}\label{eq:moweightedNs}
	 \mu_j\p=\inta{\omega}\frac{J\p_V(\omega)}{\omega^2}\omega^j=\sum_{i=1}^{N_*}S_i\p\omega_i^{\prime j},
	 \end{equation}
	 and $S_i\p=\left(\left|g_i\right|\p/\omega_i\p\right)^2$ are the Huang-Rhys parameters of the modes in the truncation. 
	 
	 For cases where $\phi(t)$ decays quickly over time, typical for spectral densities with small cut-off frequencies, we find that only a small, finite number of moments are needed for convergence. 
	 Since the lowest order moments contribute most significantly to $\mathcal{K}(\epsilon)$, the condition $\mu_j\p=\mu_j$ need only be met for a finite number of the lowest order moments. The number of moments that contribute significantly determines how many modes, $N_*$, are needed in the truncated spectral density to obtain an accurate solution. Additionally, because each mode is defined by two parameters, the Huang-Rhys parameter $S_i\p$ and energy $\omega_i\p$, defining one mode requires two simultaneous equations from the set $\mu_j\p=\mu_j$. Therefore, the $2N_*$ parameters of an $N_*$ mode truncation, $\{S_i\p\}$ and $\{\omega_i\p\}$, are found by solving the simultaneous equations $\mu_j\p=\mu_j$ for $j=1,2,3,\ldots,2N_*$. This ensures that the truncated spectral density has the same moments of the continuum spectral density that contribute significantly to the PRF. The PRF derived with $J_V\p$ is then an approximation to the full PRF described by $J_V$, and becomes exact as $N_*\to\infty$. We typically find that $N_*\le 3$ gives converged solutions, and that $N_*=1$ provides a good approximation, even for cut-off frequencies comparable to the polaron energy.  We discuss examples in Section~\ref{sec:Example}. 
	 
	 To prove the second property at $T_V>0$, we again want to identify which weighted moments $\mu_j$ are important for the mode truncation by expressing $\phi(t)$ in Eq.~\eqref{eq:phi} as a sum over the weighted moments, analogous to Eq.~\eqref{eq:phiT0}. However, $\coth(\beta_V\omega/2)$ can only be Taylor expanded for $\left|\beta_V\omega/2\right|<\pi$ and so the finite $T_V$ expansion is not possible. Instead, we can look at the $T_V\to\infty$ case for which $\coth(\beta_V\omega/2)\to 2/(\beta_V\omega)$ and we find \begin{equation}\label{eq:TVg0Expansion}
	 \lim_{T_V\to\infty}\phi(t)=\sum_{\substack{j=1\\j=\mathrm{odd}}}^{\infty}\alpha_j(t)\mu_j,
	 \end{equation}
	 where $	 \alpha_j(t)=(-1)^{\frac{j+1}{2}}t^j\left[\frac{2}{\beta_V}\frac{t}{(j+1)!}+\frac{i}{j!}\right]$. Eq.~\eqref{eq:TVg0Expansion} shows that the infinite temperature analogue of Eq.~\eqref{eq:ST0inf} will only depend on weighted moments of order $j=1,3,5,\ldots,\infty$. Eqs.~\eqref{eq:TVg0Expansion} and \eqref{eq:phiT0} imply that as $T_V$ increases from zero, the contribution of moments of even order should become smaller, until at $T_V=\infty$ only the moments of odd order contribute. The question then becomes: what temperature range is the $T_V=0$ expansion more accurate than the $T_V=\infty$ expansion? The quantity determining this is the ratio of $k_BT_V$ to the truncated mode frequencies, $\omega_i\p$. In Section~\ref{sec:Example} we will see that even for temperature ranges orders of magnitude greater than in realistic systems, the $T_V=0$ expansion is more accurate and so should always be used.
	 
	 Let us now consider examples of calculating the truncated spectral densities, using the $T_V=0$ expansion. For a single mode truncation, $N_*=1$, the truncated spectral density has the generic form: $J_V\p(\omega)=S_1\p\omega_1^{\prime 2}\delta(\omega-\omega_1\p)$. The parameters $S_1\p$ and $\omega_1\p$ are set by requiring $\mu_1\p=\mu_1$ and $\mu_2\p=\mu_2$. These lead to
	 \begin{subequations}\label{eq:singlemodeparams}
	 \begin{align}
	     &S_1\p=\lambda^2/A_V,\\
	     &\omega_1\p=A_V/\lambda,
	 \end{align}
	 \end{subequations}
	 where $\lambda$ is the reorganisation energy of the original spectral density given in Eq.~\eqref{eq:reorg} and  
	 \begin{equation}\label{eq:BathHR}
	 A_V=\mu_2=\inta{\omega} J_V(\omega),
	 \end{equation}
	 is its spectral area.
	 These are always the expressions for $S_1\p$ and $\omega_1\p$ when $N_*=1$. For a two mode truncation, $J_V\p(\omega)=\sum_{i=1}^2S_i\p\omega_i^{\prime 2}\delta(\omega-\omega_i\p)$, and setting $\mu_1\p=\mu_1$, $\mu_2\p=\mu_2$, $\mu_3\p=\mu_3$ and $\mu_4\p=\mu_4$ leads to four simultaneous equations that can be solved to find $S_1\p$, $S_2\p$, $\omega_1\p$ and $\omega_2\p$. These Huang-Rhys parameters and mode energies are the characteristic coupling strengths and mode frequencies of the bath. We note that a final criterion the spectral density $J_V$ must satisfy for the mode truncation approach to work is that its moments $\mu_j$ must be such that setting $\mu_j=\mu_j\p$ leads to positive and real values for the truncated modes. We hypothesise that this criterion is satisfied by all spectral densities
	 with finite polaron frame optical transition rates, which is certainly true for all spectral densities that we have modelled in this paper.

	 Eq.~\eqref{eq:PRFNs} is the analytic form of Eq.~\eqref{eq:PRF2} and, along with the calculation of the truncated mode properties using $\mu_j\p=\mu_j$ for $j=1,2,3,\ldots,2N_*$, is the main result of this paper. In the next section, we will use Eq.~\eqref{eq:PRFNs} to calculate optical transition rates for a variety of vibrational spectral densities and parameter regimes.

	 \section{Example calculations}\label{sec:Example}
	 For all examples in this section we will use the optical spectral density
	 \begin{equation}
	 J_O(\nu)\propto \nu^3,
	 \label{eq:JO}
	 \end{equation}
	 which arises when the Hamiltonian is derived in the multipolar gauge \cite{stokes2018master}. The form of $J_O$ does not affect the truncation method presented in this paper, although it may change the required number of modes in the truncation to obtain convergence. 
	 
	 In the following two subsections, we will compare the truncation method to an example that is numerically tractable, which will act as a benchmark for the truncation method. First we will do this at $T_V=0$ to show that a small number of modes in the truncation is sufficient, and subsequently at $T_V>0$ to show that the moments of order $j=1,2,3,\ldots,2N_*$ are the important ones even at finite $T_V$.
	 
	 The PRF in Eq.~\eqref{eq:PRF} contains three nested integrals over an infinite domain, which is difficult to solve numerically. Reliable numerical results are feasible if at least $\phi(t)$, in Eq.~\eqref{eq:phi}, is analytic and preferably if $\mathcal{K}(\epsilon)$, in Eq.~\eqref{eq:K}, is as well. Therefore, the vibrational spectral density we will use in these calculations is 
	 \begin{equation}\label{eq:JVex}
	 J_V(\omega)=S\frac{\omega^3}{\omega_c^2}\mathrm{exp}\left[-\omega/\omega_c\right],
	 \end{equation}
	 where $S=\mu_0=\inta{\omega}J_V(\omega)/\omega^2$ is the bath Huang-Rhys parameter of the bath and $\omega_c$ is the cut-off frequency. With this, both $\phi(t)$ and $\mathcal{K}(\epsilon)$ can be evaluated analytically at $T_V=0$, and only $\phi(t)$ at finite $T_V$, permitting numerical evaluation of the PRF.
	 
	 In the third subsection, we will use the truncation method to calculate the optical transition rates for vibrational spectral densities where not even $\phi(t)$ can be evaluated analytically. In these cases the truncation method is significantly easier and more reliable. 
	 
	 \subsection{$T_V=0$ numerical comparison}

\begin{figure*}[ht!]\centering
		\includegraphics[width=1\textwidth]{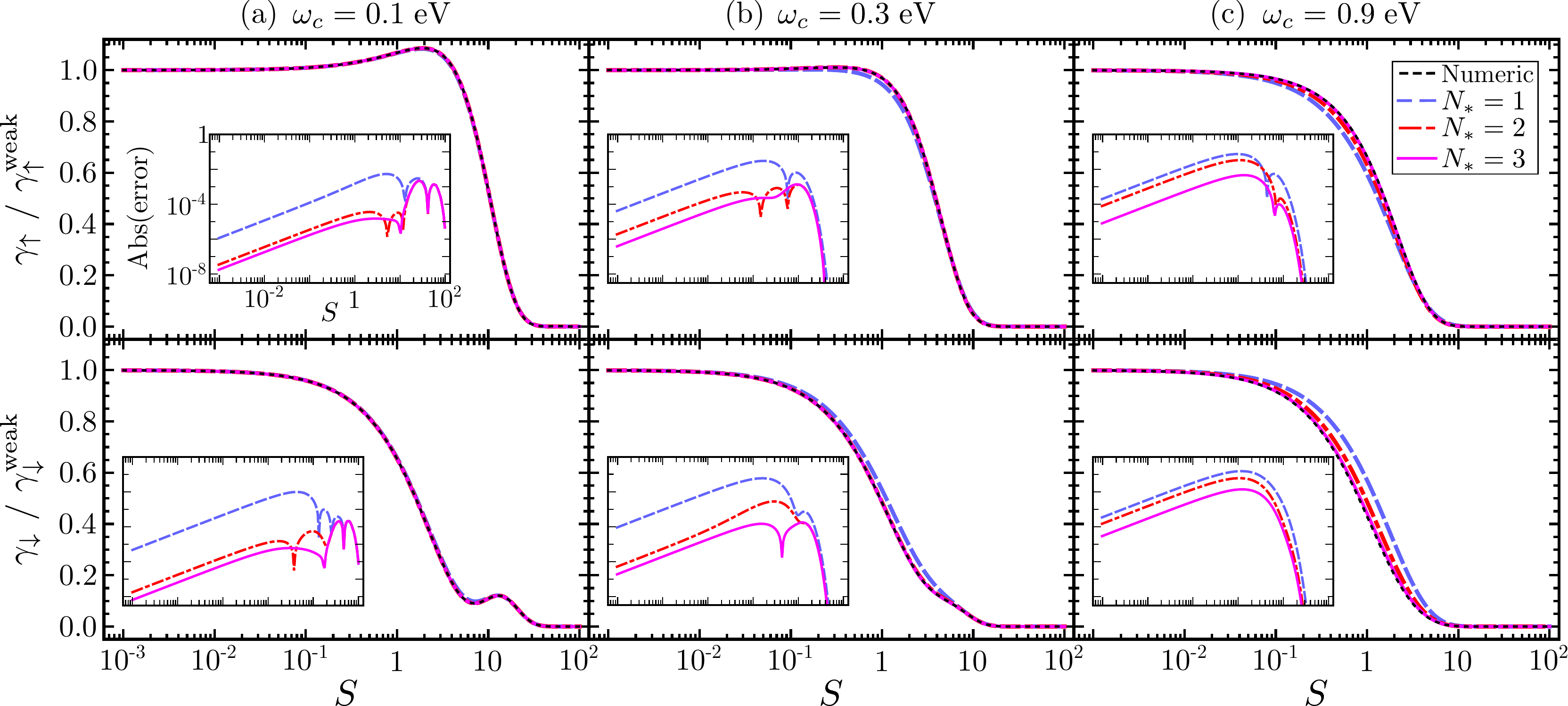}
		\caption{The normalised excitation (top row) and decay (bottom row) rates with coupling to an infinite number of vibrational modes described by the spectral density in Eq.~\eqref{eq:JVex} with $T_V=0~$K, $T_O=6000$~K and $\delta\p=1$~eV. The cut-off frequency is different in each column (a), (b) and (c). Each panel shows the rates with $N_*=1,2,3$ modes in dashed-blue, dotdashed-red and solid-purple respectively, and the numerical rates in shorter dashed-black. In the inset of each figure, which each have the same axes, we show the absolute error of each truncation compared to the numerical calculation.} 
		\label{fig:Example}
	\end{figure*}

	 In Figure~\ref{fig:Example}, we show the truncated PRF solution plotted for $N_*=1,2,3$ against the numerical solution for $T_V=0$ over a range of cut-off frequencies. The single mode truncation, with analytic PRF in Eq.~\eqref{eq:PRFsingle}, provides a surprisingly accurate approximation to the full numerical calculations over all cut-off frequencies. As expected, for larger cut-off frequencies more modes are required in the truncation to obtain numerically converged results. The non-additive behaviour of the optical and vibrational interactions is stark: increasing vibrational coupling significantly renormalises both the excitation and decay rates.
	 
	 At small cut-off frequency, in Figure~\ref{fig:Example} column (a), moderately strong vibrational coupling causes the excitation and decay rates to increase and decrease, respectively. This effect can lead to population inversion in the two-level system, which has recently been shown using the reaction coordinate mapping \cite{maguire2019environmental} and numerically using tensor networks \cite{gribben2021exact}. Excitation rates are enhanced because increasing vibrational coupling causes optical transitions to higher lying vibrational states in the excited manifold, which couple more strongly to the field owing to $J_O(\nu)\propto \nu^3$. For the same reason, when $S$ is too large the excitation rates decay exponentially because the system attempts to absorb a photon from modes that are not thermally occupied. The competition of these two effects are described by $J_A$ in Eq.~\eqref{eq:JA}. If $T_V$ was finite, then excitation could occur via spontaneous photon emission which would not require populated photon modes. 
	 
	 Turning to the decay rate in column (a), at moderate $S$, this is initially suppressed for an analogous reason that the excitation rate is enhanced: transitions occur into higher vibrational levels in the ground manifold, and so occur with an overall smaller transition energy. The greater thermal occupation of the lower energy photon modes is not sufficient to overcome the weaker coupling to them. Again, this trade off of these effects are described by $J_E$ in Eq.~\eqref{eq:JE}.  However, when $\text{round}(S)\omega_1\p\approx \delta\p$ (recall $\text{round}(S)$ is the mean number of phonons in a transition) the vibrational energy in a transition becomes comparable to the electronic splitting. For all $S$ larger than this, the dominant decay channel is no longer photon emission, but absorption. This can be seen in Figure~\ref{fig:Example}(a) where the decay rates start increasing with $S$ when $S$ is large: the decay rates become absorption processes and so are subject to the same initial enhancement as the excitation processes, described by $J_A$.
	 
	 The efficacy of a given $N_*$ truncation is not dependent on vibrational coupling strength, $S$. This is because the relative sizes of the weighted moments, $\mu_j/\mu_{j\p}$, do not scale with $S$. However, this ratio does scale with $\omega_c$ and so a large $\omega_c$ requires a greater $N_*$ for an accurate PRF solution, as seen in Figure~\ref{fig:Example}. 
	 
	 \subsection{$T_V>0$ numerical comparison}
\begin{figure}[ht!]\centering
		\includegraphics[width=0.45\textwidth]{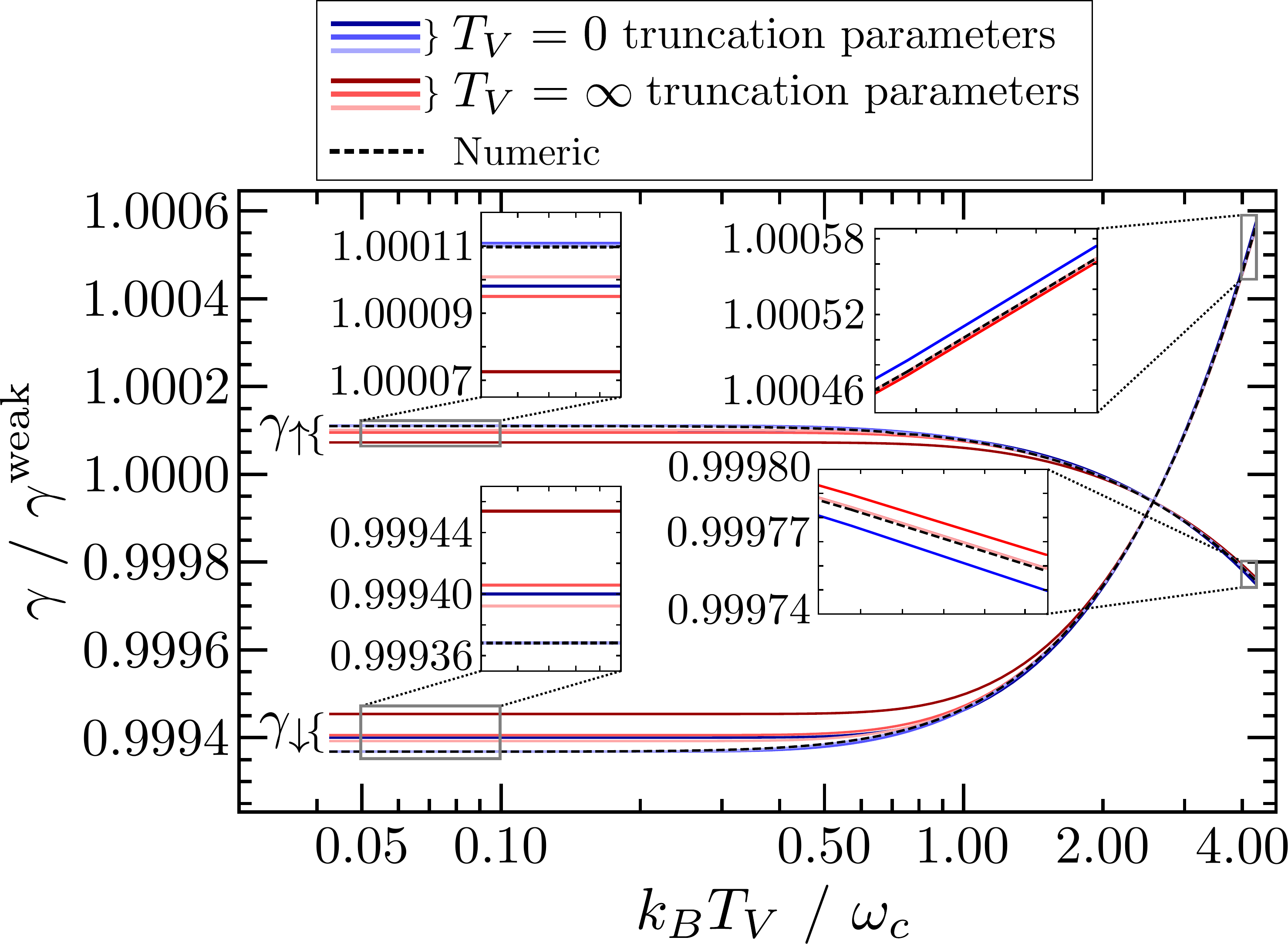}
		\caption{The vibrational temperature dependence of the optical transition rates. The vibrational spectral density is given in Eq.~\eqref{eq:JVex} with $\omega_c=0.2$~eV, and $T_O=6000$~K,  $\delta\p=1$~eV and $S=10^{-3}$. The numerical calculations are dashed-black and the finite truncation calculations using moments $j=1,2,3,\ldots,2N_*$, ($T_V=0$ expansion) in solid-blue and $j=1,3,5,\ldots,4N_*-1$, ($T_V=\infty$ expansion) in solid-red lines. For each finite truncation we plot the rates calculated using $N_*=1,2,3$ modes, with lighter colours denoting higher $N_*$. Note that all curves are produced at the same $T_V$, the solid-red and solid-blue lines differ only by the definition of the truncated mode parameters. Many curves overlap at high $T_V$ where convergence is more easily achieved. In the insets we show zoom-ins of the grey boxed regions.} 
		\label{fig:TVg0Example}
	\end{figure}

	 In Figure~\ref{fig:TVg0Example} we show the truncated mode PRF compared to the numerical solution as a function of $T_V$. The truncated PRF solution is shown for both the $T_V=0$ expansion: $j=1,2,3,\ldots,2N_*$ and $T_V=\infty$ expansion: $j=1,3,5,\ldots,4N_*-1$ for $N_*=1,2,3$ modes. At low $T_V$ for a given $N_*$ we see that the $T_V=0$ form is more accurate. Moreover, when $k_BT_V>\omega_c$, the differences in the errors of either truncation type becomes negligible. 
	 
	 As $k_BT_V/\omega_c$ increases, more vibrational levels become accessible by thermal occupation and the computational cost of calculating $A_l$ exponentially increases. For all temperature ranges in which we can feasibly calculate $A_l$ to convergence, we found that the $T_V=0$ truncation is more accurate or the same. This implies that the $T_V=\infty$ truncation is never more accurate and the weighted moments $j=1,2,3,\ldots,2N_*$ should always define the parameters of the truncated spectral density. We note that, because at $T_V>0$ only $\phi(t)$ is analytic for the full spectral density and not $\mathcal{K}(\epsilon)$, it is difficult to achieve converged numerical results \footnote{The converged series reaches higher numerical precision than the numerical approach}. Therefore, we do not show the error of each truncation compared to the numerical result.

	\subsection{Spectral densities with non-analytic $\phi(t)$}
	Finally, we show the applicability of the truncation method by using it to calculate the optical transition rates in cases for which $\phi(t)$ is non-analytic for the spectral density of interest, $J_V$. In these cases, numerical integration of the optical transition rates is difficult owing to the three nested integrals with infinite domain. The spectral densities we choose are ohmic with Gaussian and Log-normal cut-offs:
	\begin{align}
	    J_V^\text{Gauss}(\omega)&=\lambda\frac{2}{\sqrt{\pi}\omega_c}\omega\exp\left[-\left(\omega/\omega_c\right)^2\right],\\
	    J_V^\text{ln}(\omega)&=\lambda\frac{\rme^{-\frac{1}{4}}}{\sqrt{\pi}\omega_c}\omega\exp\left[-\ln^2\left(\omega/\omega_c\right)\right],
	\end{align}
	 where $\lambda$ is the reorganisation energy of the baths. Such spectral densities have been suggested for modelling the broad background of vibrational modes in bacteriochlorophyll \cite{kell2013shape,ritschel2014analytic}. For both spectral densities, $\phi(t)$ is non-analytic for $T_V>0$. For $T_V=0$, $\phi(t)$ is non-analytic for the $J_V^\text{ln}$ but analytic for $J_V^\text{Gauss}$. 
	 
\begin{figure}[ht!]\centering
		\includegraphics[width=0.45\textwidth]{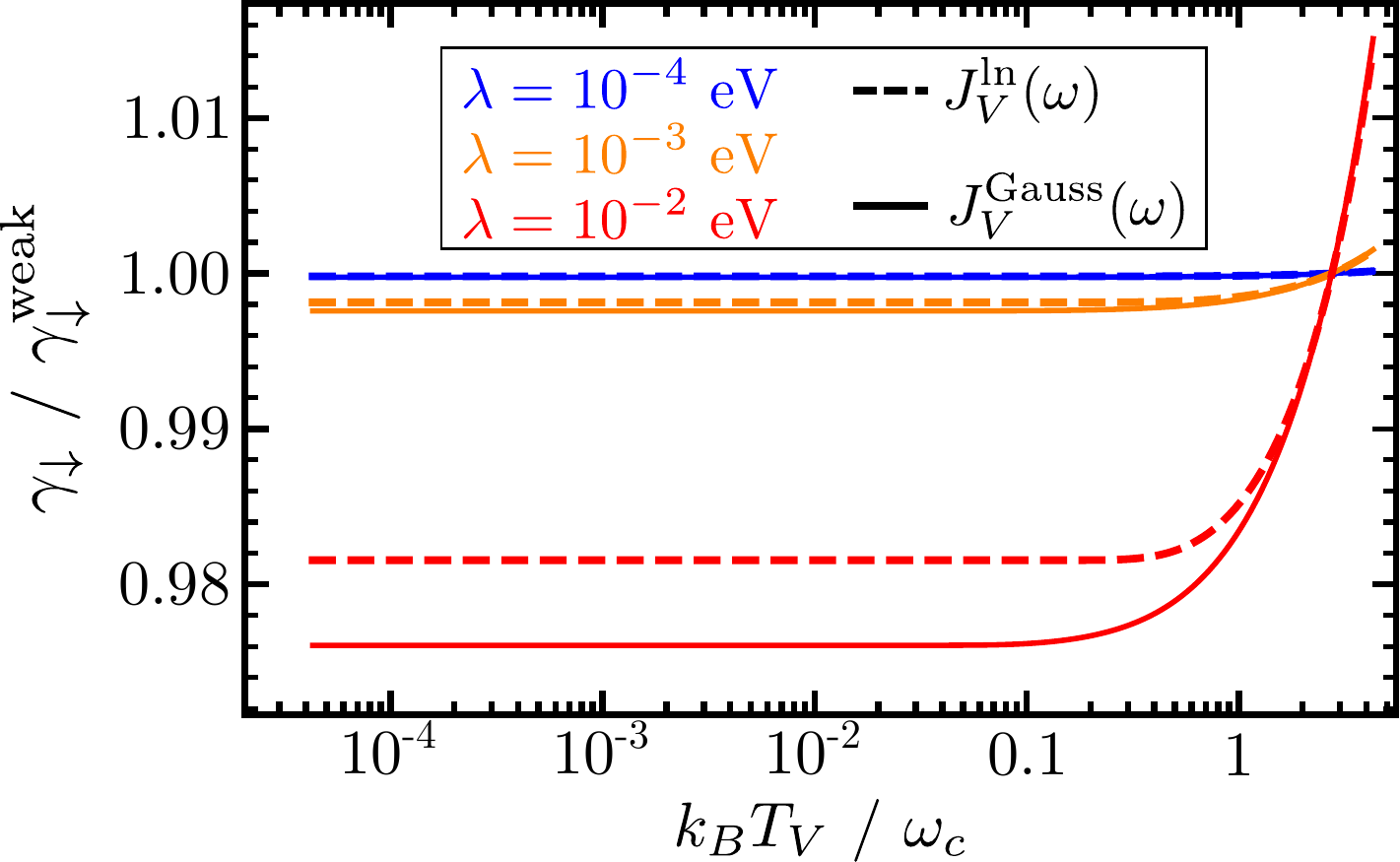}
		\caption{Decay rates for the vibrational spectral densities with Gaussian (solid lines) and log-normal (dashed lines) cut-offs as function of $T_V$, calculated with $N_*=2$ in the truncated spectral density. The reorganisation energy is varied by colour, with values $\lambda=10^{-4}$ (blue), $\lambda=10^{-3}~$eV (orange) and $\lambda=10^{-2}~$eV (red). Other parameters: $\omega_c=0.2$~eV, $T_O=6000~$K, $\delta\p=1~$eV.} 
		\label{fig:nonanalytic}
	\end{figure}

	 To gain an intuition of the comparative effects we should expect from these spectral densities we can compare the single mode parameters found with Eqs.~\eqref{eq:singlemodeparams}. $J_V^\text{Gauss}$ is represented by a mode with Huang-Rhys parameter and energy $S_1\p=\sqrt{\pi}\lambda/\omega_c$ and $\omega_1\p=\omega_c/\sqrt{\pi}$, and $J_V^\text{ln}$ by a single mode with $S_1\p=\exp[-3/4]\lambda/\omega_c$ and $\omega_1\p=\exp[3/4]\omega_c$. For a given $\lambda$ and $\omega_c$, we can therefore expect $J_V^\text{Gauss}$ to display more vibrational renormalisation (larger $S_1\p$) and be more sensitive to vibrational temperature effects (smaller $\omega_1\p$). In Figure~\ref{fig:nonanalytic} we plot the converged optical decay rates as a function of vibrational temperature at $\lambda=10^{-4}$~eV, $\lambda=10^{-3}~$eV and $\lambda=10^{-2}~$eV, calculated using the truncation method with $N_*=2$ modes (giving converged results) for both vibrational spectral densities. We have chosen relatively weak vibrational coupling strengths because we explore up to high $T_V$. As can be seen in Figure~\ref{fig:nonanalytic}, the rates with $J_V^\text{Gauss}$ are indeed more sensitive to changes in both $\lambda$ and $T_V$. We note that the intuition gained from looking at the single mode parameters is maintained in spite of the calculations having been performed with $N_*=2$. 
	
	 Although the truncation method works easily for spectral densities where $\phi(t)$ cannot be evaluated numerically, this method cannot handle vibrational spectral densities with any moments $\mu_j$ for $j=1,2,3,\ldots,\infty$ that diverge. For example, this includes the Drude-Lorentz, overdamped and underdamped spectral densities commonly used to model vibrational environments \cite{maguire2019environmental}. However, this shortcoming is one of the polaron transformation: for these spectral densities $\phi(t)$ is itself divergent.
	 
	 \section{Conclusion}
	 We have developed an analytic form of the polaron rate function, Eq.~\eqref{eq:PRF2}, that ubiquitously appears in polaron theory when deriving optical transition rates. The expression, in Eq.~\eqref{eq:PRFNs}, relies on a truncated spectral density which emulates the effects of the actual spectral density but with a finite number of modes. The expression converges faster for smaller cut-off frequencies where the phonon propagator decays quickly but still offers accurate calculations far outside of this regime. The required truncated mode expansion is different for zero or infinite vibrational temperature. However, we have shown that even for $k_BT_V$ much larger than the vibrational mode energies, the zero temperature expansion converges for fewer modes in the truncation. This result will be useful whenever polaron theory is applied to systems interacting with optical baths so that the non-additive physics of systems weakly coupled to optical baths and strongly coupled to vibrational baths is captured. This is particularly true when one wishes to use vibrational spectral densities for which $\phi(t)$ cannot be calculated analytically where the only alternative is a difficult numerical integration of the PRF.
	 
	 As well as providing an easily computed solution, the finite truncated spectral density $J_V\p$ provides physical insight into the full spectral density, $J_V$, one wishes to model. When a single mode truncation is effective, the vibrational interaction is equivalent to a single mode of Huang-Rhys parameter $S_1\p=\lambda^2/A_V$ and energy $\omega_1\p=A_V/\lambda$, which easily allows one to determine a characteristic coupling strength and energy scale for the full vibrational bath. This is then useful, for example, for determining when finite vibrational temperature effects become important: this is when $k_BT_V\sim\omega_1\p$. The same is true for an $N_*>1$ truncation, only that the bath has more than one characteristic energy and coupling strength.

	 \begin{acknowledgements}
We would like to thank Dominic Gribben, Jonathan Keeling, Ahsan Nazir, Jake Iles-Smith and Peter Kirton for insightful discussions. D. M. R. acknowledge studentship funding from EPSRC (EP/L015110/1). B. W. L. and E. M. G. acknowledge support from EPSRC (grants EP/T014032/1 and  EP/T01377X/1).
\end{acknowledgements}

\bibliographystyle{myunsrt.bst}	 
\bibliography{paperbib}

\end{document}